\documentclass[sn-mathphys]{sn-jnl}
\usepackage{amsmath}
\usepackage{amsfonts}
\usepackage{amssymb}
\usepackage[title]{appendix}
\usepackage{graphicx}
\usepackage{sidecap}
\usepackage{rotating}
\usepackage{caption}
\usepackage{subcaption}
\usepackage{bm}% bold math
\usepackage{graphicx}
\newcommand{\Rmn}[1]{\uppercase\expandafter{\romannumeral #1}}

\allowdisplaybreaks
\makeindex

\begin{document}
\title{Zero-point energy of a trapped ultracold Fermi gas at unitarity: squeezing the Heisenberg uncertainty principle and suppressing the Pauli principle to produce a superfluid state}
\author{\fnm{D. K.} \sur{Watson}} \email{dwatson@ou.edu}
\affil{\orgdiv{Homer L. Dodge Department of Physics and Astronomy},
\orgname{University of Oklahoma},
\orgaddress{\city{Norman}, \postcode{73019}, \state{Oklahoma}, \country{USA}}}

\date{\today}

\abstract
{The zero-point energy of a trapped ultracold Fermi gas at unitarity
is investigated in relation
to the combined effects of the Heisenberg uncertainty principle and the
Pauli principle.  This lowest allowed quantum state is a superfluid state
which has been studied extensively both experimentally and theoretically.
The method used for the current investigation is based on a recent series of
papers that proposed microscopic dynamics based on normal modes to describe
superfluidity instead of real-space Cooper pairs. This approach yielded
excellent agreement with experimental data for multiple properties and 
allowed the microscopic behavior underlying these results as well as the
basis of universal behavior to be analyzed in
detail using the group theoretic basis of this general N-body approach.
This microscopic picture is now used to illucidate the roles
played by the uncertainty principle and the Pauli principle in determining
the energy and character of the lowest allowed quantum state
including the squeezed character of this superfluid state and the suppression
of the Pauli principle.}
 
\bigskip


\smallskip

\noindent ORCID iD:  D.K. Watson https://orcid.org/0000-0001-8678-7745

\smallskip

%\section{Keywords}: zero point energy, Pauli principle, collective phenomena,
%Heisenberg uncertainty principle

\maketitle

\section{Introduction}
Zero-point energy is defined as the lowest possible energy that a
quantum system of particles may have consistent with the Heisenberg
uncertainty principle. Even at absolute zero, quantum particles retain
some motion, continuously fluctuating about values of
the particles' positions and momentums as a consequence of their
wave-like nature.  This residual energy is known to result
in measureable physical
effects. For example, due to its large zero-point energy and low mass,
helium remains a liquid down to absolute zero
at standard atmospheric pressure. Other physical effects include
spontaneous emission, the Lamb shift, the Casimir effect,
and the magnetic  moment of the electron.
Zero-point energy is an important concept
in multiple fields of physics affecting the discrepancy between
the observed and calculated vacuum energy in cosmology\cite{maggiore},
the validity of supersymmetry in high energy physics\cite{zumino,nielsen}, 
the fluctuating fields of quantum field theory\cite{milonni} and offering a
possible source of dark energy\cite{maggiore,lehnert}.  Systems whose zero-point energy
signatures have recently been studied include complex
molecules\cite{richard}, nanocrystals\cite{duan}, optomechanical
systems\cite{lecocq,chegnizadeh}, superconducting materials\cite{errea},
and redshifted light from distant
galaxies\cite{setterfield}.

The ground state of ultracold Fermi gases has
been studied extensively both experimentally and theoretically
across the BCS to unitarity transition, i.e. from weak
interactions limiting to the independent particle case to the strongly
interacting unitary regime\cite{hulet1,jin1,zwierlein2,jochim1,thomas1,salomon1,jin2,zwierlein1,grimm1}. Theoretical descriptions of this transition 
can offer a test of the microscopic dynamics that underlie
the evolution of the zero-point energy, i.e.
the ground state energy of the system across this transition.
As will be demonstated in this study,
both the uncertainty principle and the Pauli principle contribute to
the changes in dynamics that are responsible for the character and energy
of this lowest state and the emergence of
superfluid behavior.

\smallskip

\subsection{Background}
Typically the ground state energy at zero temperature
has been viewed theoretically as simply the lowest
energy state of the relevant Hamiltonian chosen to describe this
ultracold Fermi gas. In this work, I will view the ground state energy from
a different perspective, that of the zero-point energy 
of a many-body system i.e. the minimum energy 
required to satisfy both the Heisenberg uncertainty principle and the Pauli
principle for this system of identical fermions at zero temperature.

\smallskip
\noindent{\bf{\small{ 1) The Heisenberg uncertainty principle. -}}} The zero-point energy is fundamentally a consequence of the Heisenberg
uncertainty principle. Specifically, the uncertainty principle forbids
complementary variables such as position and momentum (or energy and time)
to be precisely specified by any quantum state. Therefore, the lowest energy
state must have a distribution in position and momentum that satisfies the
uncertainty principle, and thus its energy can never be zero even at a
temperature of absolute
zero. The product of the uncertainties for each particle in each
dimension must obey:

\begin{equation}
\Delta x \Delta p \ge \hbar/2
\end{equation}

For a three-dimensional quantum harmonic oscillator, the product of the
uncertainties in each dimension in
the lowest state saturates this lower limit of the uncertainty principle
with the two uncertainties
equal, i.e. with the same amount of uncertainty residing in the position
of the particle and its momentum.  This yields a zero-point energy
of $3\hbar\omega_{ho}/2$ for each particle in this lowest state. This type of
minimum uncertainty state with symmetric uncertainties is sometimes
referred to as a coherent state.  Higher states of the simple harmonic
oscillator have larger values of both $\Delta x$ and $\Delta p$ as the energy
of the state increases and the spatial extent of the wave function expands.
For a many-body system of particles, each particle must individually satisfy
the Heisenberg uncertainty principle.

Quantum states in which the uncertainty in one complementary variable has
been reduced below the symmetric limit while the uncertainty in the
other variable has increased such that the Heisenberg uncertainty principle is
satisfied are called squeezed states.  Such states have been known
since the earliest days of quantum mechanics, have been used for decades
in precision measurements in spectroscopy and
interferometry including tests of fundamental physics, improvements in atomic
clocks\cite{caves,bigelow,grangier,gerbier,huang,zwierlein3,schleier-smith,ye} 
and improving the detection capabilities of gravitational-wave
detectors\cite{abadie,aasi}.

\smallskip
\noindent{\bf{\small{ 2) The Pauli principle. -}}}The occupation of these
states for
independent, identical fermions in a harmonic well is, of course, determined by the Pauli
principle; thus the lowest energy, i.e. the zero-point energy of an N-body
system of identical fermions is controlled by the Pauli principle. The Pauli
principle forbids two or more identical fermions, i.e. with all quantum
numbers the same, from occupying the same quantum
state. If two otherwise identical fermions have opposite spins, they may reside
in the same quantum state, filling the allowed states two-by-two.
This result can be compared to the situation for a system of identical
non-interacting bosons in a three-dimensional harmonic trap
all of which can occupy the lowest quantum state, so
all the bosons in a N-body system can be in the minimum uncertainty
state yielding a zero-point energy of $3N/2\hbar\omega_{ho}$. Thus we expect the
zero-point energy of a system of identical non-interacting
bosons to be much lower than
the zero-point energy of a N-body system of identical non-interacting
fermions each of which must
obey the Pauli principle as they fill the available states.

\smallskip
\noindent{\bf{\small{ 3) A normal mode picture. -}}}
Recent studies have investigated the superfluidity of ultracold Fermi 
gases\cite{inguscio,randeria,zwerger,strinati} using
a perturbation formalism called 
symmetry-invariant perturbation theory, SPT.
This approach, which is a first-principles, general, many-body method with
no adjustable parameters,
employs group theory and graphical techniques to solve each perturbation
order, in principle exactly, rather than using
a basis set or numerical methods. The first-order equation, which is
harmonic, has been solved exactly through the determination of
the group theoretic normal modes with $N$ as a parameter\cite{FGpaper,paperI,JMPpaper}. 
 The Pauli principle is applied using an adiabatic transition
from independent particles to an interacting 
regime\cite{harmoniumpra,prl}  without explicit 
antisymmetrization of the many-body wave function thus circumventing the
intensive numerical work typically required.
Despite no explicit real-space two body pairing,
 first-order 
results yielded close agreement
with experimental data for multiple properties
without higher order terms both in the
strongly-interacting unitarity regime as well as for weaker interactions
from the BCS regime to unitarity\cite{prl,jlow,emergence,jlow2}.
These properties included ground state energies\cite{prl},
critical temperatures\cite{jlow}, 
excitation frequencies\cite{jlow,prafreq},  
thermodynamic entropies and energies\cite{jlow,emergence}, as well as
the compressibility\cite{jlow2} and the lambda transition in the 
specific heat\cite{emergence},  well known signatures of the onset of
superfluidity. These results were obtained with no adjustable parameters
and with seconds of desktop computer time due to the use of symmetry in
the group theoretic approach and the enforcement of the Pauli principle
'on paper' which avoids the explicit determination of a many-body
antisymmetric wave function.
(A brief discussion of these results and an
explanation of the underlying microscopic basis can be found in 
Supplemental Information Section I. A-D. in Ref.~\cite{EPL}. For
more detail, see
Refs.~\cite{jlow,emergence,prafreq}.)

These results suggested the possibility 
that normal modes might offer an alternative microscopic basis
for superfluid behavior\cite{EPL}
%\cite{annphys,prafreq} 
differing from the conventional view 
that some fermions form loosely bound pairs 
that condense into a macroscopic
occupation of the lowest 
state\cite{hulet1,jin1,zwierlein2,jochim1,thomas1,salomon1,jin2,zwierlein1,grimm1,leggett1,leggett2,eagles,nozieres,salasnich}.  In addition to
producing good agreement
with experimental data of multiple properties, normal mode dynamics
 have suggested an interesting microscopic explanation 
for the universal
behavior at unitarity\cite{prafreq,annphys}.

Offering an alternative route 
to a phase coherent macroscopic wave function using normal modes necessarily
draws attention to the importance of inter-pair correlations 
which are due to the Pauli principle and which have always been recognized as 
crucial to accurately describe
superconductivity/superfluidity\cite{combescot1,combescot2,combescot3}.
Based on these findings, the goal of this study is to 
understand the roles that the Heisenberg uncertainty principle and the Pauli
principle play in producing a superfluid state. This approach will utilize
the group
theoretic expression for the energy through first order and the restrictions
imposed by these two fundamental principles to analyze the effects that combine
to produce and support superfluid behavior.

\smallskip
\noindent{\bf{\small{ 4) Reimagining superfluidity. -}}}
The perspective of fermionic superfluidity used in this study of
zero-point energy is an extension of previous
work that reimagined the microscopic basis of
superfluidity\cite{EPL}.  This reimagination adheres
 closely  to the early tenets of superconductivity/superfluidity\cite{bcs}
which assumed two-body pairing
solely in momentum space, not in real space, and was used to reinterpret 
several interrelated 
phenomena including Cooper pairs, the Fermi sea, and Pauli blocking.
The Pauli principle, in its recently revealed role in collective motion,
played a critical role in the proposed dynamics selecting the 
allowed normal modes.

%\noindent {\bf{\large{ Background. -}}}
The Pauli principle is known to dominate the inter-pair interactions in the 
BCS ansatz\cite{combescot1,combescot2,combescot3}, 
and is instrumental to 
producing key properties of superfluidity/superconductivity
including an excitation gap, the rigidity of the
wave function responsible for the Meissner effect, and the 
absence of resistance to current flow.
It is interesting that early work did not assume  
two-body pairing in real space.
The highly successful BCS theory proposed in 1957\cite{bcs} 
assumes the pairing of fermions only in momentum space
with $+k$ and $-k$ values.
As the 1957 paper states, 
the BCS wave function describes 
the ``coherence of large numbers of electrons,'' 
but does not propose fermion pairs localized into
pseudomolecules that transition as in Bose-Einstein
condensation\cite{bcs}. As suggested by London in
1950, a superconductor is a ``quantum structure on a macroscopic scale...
a kind of solidification or condensation of the average momentum distribution'' of the 
electrons. ``It would not be due to distinct electrons at separate
places having the same momentum'', but ``it would arise from wave packets of
wide extension in space assigning the same local momentum to the entire
superconductor''\cite{london1}.  These early concepts of 
superfluidity/superconductivity as well as the seminal properties: 
 momentum space pairing, the long-range order over macroscopic distances,
 the wave function ``rigidity'', and the gap in the excitation 
spectrum\cite{london2,schrieffer} are
naturally present in a normal mode picture of superfluidity.

In this paper, I will use the formula for the ground state, i.e. zero-point
energy of a Fermi
gas from the first order SPT solution to analyze the contributions from both
the Heisenberg uncertainty principle and the Pauli principle to this energy.
Since the Pauli principle has been applied 'on paper' through an adiabatic
transition to the interacting case, this formula clearly shows the
contribution of the Pauli principle to this energy.
I will separate out this Pauli contribution and define the remainder as
the 'non-Pauli' contribution to the zero-point energy, due only to
the uncertainty principle i.e. the lowest energy that would be possible if
all the particles could reside in the lowest state. I will
compare these separate contributions to the corresponding contributions
for 
independent, non-interacting identical fermions in a three dimensional well for
which the Pauli principle
contribution is, of course, well known.  These two fundamental
principles will be shown
to work together to create a very low energy state with a condensed i.e.
squeezed
value of the momentum and an extended state in position space.
This state lies significantly lower than the next higher normal mode
state creating a gap
that supports superfluid behavior.

\section{Symmetry-Invariant Perturbation Theory: a group theoretic
and graphical approach to the general N-body problem}

\smallskip
\noindent{\bf{\small{ 1) Overview. -}}}
Symmetry-invariant perturbation theory is a non-numerical approach to
many-body problems.  It uses symmetry rather than
intensive numerical methods to describe
many-body systems, and thus can offer clear physical insight into
the underlying 
dynamics when higher order terms are small.
The perturbation parameter 
is the inverse dimensionality of space. 
Originally developed by t'Hooft in quantum chromodynamics\cite{t'hooft}, 
 and subsequently used by Wilson\cite{kenwilson}
in condensed matter to calculate
 critical exponents for $D=3$ phase transitions
 starting from the $D=4$ exact values, $1/D$ or $1/N$ expansions have now
 been developed to study physical systems  in multiple fields
 of physics from atomic and
 molecular\cite{survey1,survey3,many1,many2,many3,loeser,atomic1,atomic2,atomic3,atomic4,atomic5,atomic6,atomic7,atomic8,atomic9,atomic10,highfields2,HatomEM2,quasi2} 
and condensed matter
\cite{kenwilson,condensed1,condensed2,condensed3}, 
to quantum field 
theory\cite{whitten,quantumfield1,quantumfield2,quantumfield3,quantumfield4,
quantumfield5,survey6}.

Initially, the SPT formalism was developed to handle
the large ultracold ensembles of bosons being studied in atomic
physics/condensed matter communities\cite{FGpaper,paperI,JMPpaper,energy,laingdensity,test,toth}. More recently, it was extended 
to ultracold Fermi gases\cite{prl,jlow,emergence} which are subject to 
Pauli constraints\cite{harmoniumpra,prl,emergence,partition}. 
Currently, this method is 
formulated through first order for three dimensional $L=0$ 
systems with spherically-symmetric confinement potentials
and general interaction potentials. 
The SPT approach uses symmetry to tackle the $N$-scaling
problem\cite{FGpaper,paperI,JMPpaper,liu,montina2008}, 
rearranging the work required for an
exact solution so the exponential scaling depends
on the order of the series, not the value of $N$ as is the case in conventional
many-body approaches. 
To take advantage of maximal symmetry, a perturbation series is formulated
about a large-dimension structure whose point group is
isomorphic to the symmetric group $S_N$\,, 
then evaluated for $D=3$.
This approach enables the extraction of the
 work at each order that scales exponentially as a pure mathematical
problem  (cf. the Wigner-Eckart theorem)\cite{rearrangeprl,complexity}.
In principle, this extracted problem can be solved exactly
using group theoretic 
methods, and saved\cite{epaps}, significantly reducing the  
numerical cost and underpinning the robustness of this non-numerical
approach.

Unlike conventional perturbation approaches, the perturbation in SPT does not 
involve the strength of the interaction and so it can be applied to
strongly interacting systems such as the unitary regime. This many-body
approach, however,
does not offer a mechanism for the transition to diatomic molecules
in the BEC regime.  In principle, the 
BEC regime could be described by including higher-order terms 
although the number of terms required would probably  undermine
any physical insight and going to higher order is now
exponentially difficult.

Since even the lowest order contains
beyond-mean-field effects, excellent first-order results
have been obtained\cite{prl,emergence,energy} as seen in 
earlier dimensional approaches\cite{loeser,herschbach1,herschbach2,kais1,kais2}.
The current formalism has been extensively tested on a fully-interacting
 model problem of harmonically-confined, harmonically-interacting 
particles\cite{harmoniumpra,test,toth,partition}. 
Agreement of ten or more digits was obtained
between the SPT wave function and the exact wave function
obtained independently
verifying this many-body formalism\cite{test} 
and the analytic forms for the group-theoretic, $N$-body normal modes.

\smallskip
\noindent {\bf{\small{ 2) The SPT formalism. -}}}
In $D$ dimensional Cartesian coordinates,
the $N$-body Schr\"odinger equation  can be written:
\begin{equation} \label{eq:generalH} 
H \Psi  =  \left[ \sum\limits_{i=1}^{N} h_{i} +
\sum_{i=1}^{N-1}\sum\limits_{j=i+1}^{N} g_{ij} \right] \Psi = E
\Psi \,,  
\end{equation} 
\begin{equation} \label{eq:generalH1} 
\begin{array}{rcl}
h_{i} & = & -\frac{\hbar^2}{2
m_{i}}\sum\limits_{\nu=1}^{D}\frac{\partial^2}{\partial
x_{i\nu}^2} +
V_{\mathtt{conf}}\left(\sqrt{\sum\nolimits_{\nu=1}^{D}x_{i\nu}^2}\right)
\,,  \\
g_{ij} & = & V_{\mathtt{int}}\left(\sqrt{\sum\nolimits_{\nu=1}^{D}\left(x_{i\nu}-x_{j\nu}
\right)^2}\right),  
\end{array}
\end{equation}
\noindent where $h_{i}$ represents a single-particle Hamiltonian,
$ V_{\mathtt{int}}$ denotes a general two-body interaction potential,
 $x_{i\nu}$ refers to the $\nu^{th}$ Cartesian component
of the $i^{th}$ particle, and 
$V_{\mathtt{conf}}$ is a spherically-symmetric confining 
potential\cite{FGpaper,paperI,JMPpaper}. 
The Hamiltonian is transformed to internal 
coordinates, $r_i$ and $\gamma_{ij}$,  where
$r_i=\sqrt{\sum_{\nu=1}^{D} x_{i\nu}^2}\,, \;\;\; (1 \le i \le
N)\,,
\;\;\;$ %\mbox{and} %\;\;\;
are the $N$ $D$-dimensional scalar radii
and 
$\gamma_{ij}=cos(\theta_{ij}) \\
=\left(\sum_{\nu=1}^{D}x_{i\nu}x_{j\nu}\right) / r_i r_j\,,\;(1 \le i < j \le N)$, 
are the $N(N-1)/2$ cosines of the angles between the radial
vectors.

The first-order derivatives are removed using a
similarity transformation\cite{avery}, and
the large-dimension limit of the Schr\"odinger equation is
regularized by a scale factor.
Substituting the scaled variables
and defining the perturbation parameter as $\delta=1/D$ gives:
\begin{equation} \label{eq:scaleH_BEC}
\bar{H} \Phi =
\left(\delta^2\bar{T}+\bar{U}+\bar{V}_{\mathtt{conf}}+\bar{V}_{\mathtt{int}}\right)\Phi = \bar{E}\, \Phi\,.
\end{equation}

\begin{eqnarray}
\label{eq:Ubar_BEC1}
\bar{T} &=& \sum\limits_{i=1}^{N} \Bigl( -\frac{1}{2}\frac{\partial^2}
{{\partial \bar{r}_i}^2} 
 - \frac{1}{2 \bar{r}_i^2}\sum\limits_{j\not=i}\sum\limits_{k\not=i}
\frac{\partial}{\partial\gamma_{ij}}(\gamma_{jk}-\gamma_{ij}\gamma_{ik})
\frac{\partial}{\partial\gamma_{ik}} \Bigr) \nonumber \\
\bar{U}&=&\sum\limits_{i=1}^{N}\left(\frac{\delta^2N(N-2)+(1-\delta(N+1))^2 \left(\frac{\Gamma^{(i)}}{\Gamma}\right)}{8 \bar{r}_i^2}\right) \, \nonumber
\end{eqnarray}
\begin{equation}
\bar{V}_{\mathtt{conf}}=\sum\limits_{i=1}^{N}\frac{1}{2}\bar{r}_i^2,\,\,\,\,
\bar{V}_{\mathtt{int}}=  \frac{\bar{V}_{0}}{1-3b'\delta}
\sum\limits_{i=1}^{N-1}\sum\limits_{j=i+1}^{N}
%\left(
1-\tanh\Theta_{i,j} \nonumber
%\right) 
\end{equation}
\noindent The Gramian determinant $\Gamma$ contains elements $\gamma_{ij}$ (see Appendix D in Ref~\cite{FGpaper}), and the
$\Gamma^{(i)}$ determinant excludes the row
and column of the $i^{th}$ particle.  

Defining
$\Theta_{ij}=\frac{\bar{c}_0}{1-3\delta}
\left(\frac{\bar{r}_{ij}}{\sqrt{2}}-\bar{\alpha}-3\delta\left(\bar{R}-\bar{\alpha} \right) \right)
,$
with 
$\bar{r}_{ij}={\sqrt{{\bar{r}_i}^2+
{\bar{r}_j}^2-2\bar{r}_i\bar{r}_j\gamma_{ij}}}\,$
%\end{equation} 
the interatomic separation, $\bar{R}$ the range of the square-well potential,
and $\bar{\alpha}$  a constant which softens the
potential as $D \rightarrow \infty$,
the form of $\bar{V}_{\mathtt{int}}$ 
reduces to a square well at $D=3$ and is differentiable away
from $D=3$ permitting the dimensional analysis\cite{FGpaper,energy}.
The constant $b'$ is 
selected so at unitarity with $\bar{V}_0 = 1.0$, the
scattering length is infinite.
To reach the weaker 
interactions of the BCS regime, $\bar{V}_0$ is scaled to smaller
values.
The range of the square-well potential $\bar{R}<<\bar{a}_{ho}$ 
and is systematically reduced to extrapolate 
to zero-range interaction.

When $D\to\infty$, the second derivative terms drop out
 producing a static problem at zeroth order 
with an effective potential, $\bar{V}_{\mathtt{eff}}$:
\begin{equation}\label{eq:veff_BEC}
\bar{V}_{\mathtt{eff}} =
\sum\limits_{i=1}^{N}
\bar{U}(\bar{r}_i;\delta)
+\bar{V}_{\mathtt{conf}}(\bar{r}_i;\delta)
 +\sum\limits_{i=1}^{N-1}\sum\limits_{j=i+1}^{N}
\bar{V}_{\mathtt{int}}(\bar{r}_i,\gamma_{ij};\delta). \nonumber
\end{equation}
\noindent The minimum of this effective potential
is an infinite-dimensional maximally-symmetric
structure 
with all $\bar{r}_i$ and  $\gamma_{ij}$ equal: for $D\to\infty$, 
$\bar{r}_{i}=\bar{r}_{\infty}(1\le i \le N)$ and
$\gamma_{ij}={\gamma}_\infty(1\le i < j \le N)$.
Two minimum conditions: 
$\left(\frac{\partial \bar{V}_\text{eff}}
{\partial\bar{r}_{i}}\right)\Biggr|_{\infty}=0, 
\left(\frac{\partial \bar{V}_\text{eff}}
{\partial\gamma_{ij}}\right)\Biggr|_{\infty}=0$
yield two equations in $\bar{r}_\infty$ and $\gamma_\infty$:
\,$\bar{r}_\infty=\frac{1}{\sqrt{2} \sqrt{1+(N-1) \gamma _{\infty }}}$
 and
$\frac{\gamma _{\infty } \left(2+(N-2) \gamma _{\infty
   }\right)}{\left(1-\gamma _{\infty }\right){}^{3/2}
   \sqrt{1+(N-1) \gamma _{\infty }}}
   +
   \bar{V}_0\,\text{sech}^2\left(\Theta _{\infty}\right)\Theta _{\infty }'
   =
   0\,.$
\noindent Expanding about the minimum ($r_\infty, \gamma_\infty$): 
$\bar{r}_{i} = \bar{r}_{\infty}+\delta^{1/2}\bar{r}'_{i}$
and $\gamma_{ij} =
{\gamma}_{\infty}+\delta^{1/2}{\gamma}'_{ij}$, sets up a
power series in $\delta^{1/2}$.

\smallskip
\noindent{\bf{\small{ 3) The FG method. -}}}
The first-order, $\delta=1/D$, equation is  harmonic
and is solved exactly using group theory to obtain
 the $N$-body normal modes\cite{FGpaper,paperI,JMPpaper}.
 The first-order  Hamiltonian, $\bar{H}_1$,
%$\widehat{H}_1$
 is defined in terms of the constant matrices, 
 ${\bf F}$ and $\bm{G}$ composed of potential terms and kinetic energy
 terms respectively, evaluated at the large
dimension limit:
%(see Appendix~\ref{app:Review}):
\begin{equation} \label{eq:Gham}
\bar{H}_1=-\frac{1}{2} {\partial_{\bar{y}'}}^{T} {\bm G}
{\partial_{\bar{y}'}} + \frac{1}{2} \bar{\bm{y}}^{\prime T} {\bm
F} {{\bar{\bm{y}}'}} + v_o \,,
\end{equation}
\noindent with $v_o$ a constant\cite{FGpaper}.
The FG matrix method\cite{dcw}, extensively used in
molecular physics, is utilized to obtain the normal-mode
frequencies\cite{FGpaper} and coordinates\cite{paperI}. (Appendix A 
in Ref.~\cite{FGpaper} contains a brief summary.) 
%but a brief summary is given in Appendix~\ref{app:Review} of this paper.
Only five distinct frequencies, $\bar{\omega}$, are obtained regardless of
the value of $N$. 
%$P \equiv
This large degeneracy reflects the
% (see Ref.~\cite{hamermesh} and Appendix~\ref{app:Char}),
very high degree of symmetry in
the $\bm{F}$ and $\bm{G}$
%and $\bm{FG}$ 
matrices which are evaluated for the $D\to\infty$,
maximally-symmetric structure which has a single value for all  
$\bar{r}_\infty$ and 
 ${\gamma}_{\infty}$. These matrices are thus
invariant under the $N!$ particle interchanges 
of $S_N$ and do not connect
subspaces belonging to different irreducible representations (irreps) of 
$S_N$\cite{hamermesh,WDC}. Consequently, 
%from Eqs.~(\ref{eq:qytapp}) and (\ref{eq:FGitapp})
the normal coordinates transform under irreps of $S_N$.

The five irreps include: a one-dimensional radial
and a one-dimensional angular irrep 
 both labelled by the partition $[N]$,
 an $N-1$-dimensional radial and an $N-1$-dimensional angular irrep
 both labelled by the 
partition $[N-1, \hspace{1ex} 1]$,
and an angular irrep of dimension $N(N-3)/2$  labelled
by $[N-2, \hspace{1ex} 2]$.
These irreps are denoted by shorthand labels:
 ${\bf 0}^-$, ${\bf 0}^+$, ${\bf 1}^-$, 
${\bf 1}^+$, and ${\bf 2}$ respectively,
(see Refs.~\cite{paperI,JMPpaper}).
The character of these five modes has been analyzed as a function of
$N$ in Ref.~\cite{annphys}.
For $N$ large, 
 the single normal mode of type ${\bm 0^+}$  
has center of mass motion;
the single ${\bm 0^-}$ mode is a 
breathing motion;
the $N-1$  type ${\bm 1^+}$ modes exhibit particle-hole/single-particle
angular excitation behavior;  the $N-1$
   type ${\bm 1^-}$  modes display particle-hole i.e.
single-particle radial excitation behavior;  and
the $N(N-3)/2$  
 type ${\bf 2}$ modes correspond to phonon compressional modes.

A symmetry coordinate vector, $S$, is defined:
\begin{equation}\label{eq:trial}
\bm{S} = \left( \begin{array}{l} {\bm{S}}_{\bar{\bm{r}}'}^{[N]} \\
{\bm{S}}_{\overline{\bm{\gamma}}'}^{[N]} \\
{\bm{S}}_{\bar{\bm{r}}'}^{[N-1, \hspace{1ex} 1]} \\
{\bm{S}}_{\overline{\bm{\gamma}}'}^{[N-1, \hspace{1ex} 1]} \\
{\bm{S}}_{\overline{\bm{\gamma}}'}^{[N-2, \hspace{1ex} 2]}
\end{array} \right) =
\left( \begin{array}{l}W_{\bar{\bm{r}}'}^{[N]} \, \bar{\bm{r}}' \\ 
W_{\overline{\bm{\gamma}}'}^{[N]} \, \overline{\bm{\gamma}}' \\
W_{\bar{\bm{r}}'}^{[N-1, \hspace{1ex} 1]} \bar{\bm{r}}'\\
W_{\overline{\bm{\gamma}}'}^{[N-1, \hspace{1ex} 1]} \,
\overline{\bm{\gamma}}' \\
W_{\overline{\bm{\gamma}}'}^{[N-2, \hspace{1ex} 2]} \,
\overline{\bm{\gamma}}' \end{array} \right) \,,
\end{equation}
where  $W_{\bar{\bm{r}}'}^{[\alpha]}$ 
and $ W_{\bar{\bm{\gamma}}'}^{[\alpha]}$ are transformation matrices
determined using the theory of group characters 
to decompose
$\bar{\bm{r}}'$ and $\overline{\bm{\gamma}}'$ into basis functions that 
transform under the five irreps of $S_N$\cite{paperI}. 
The $\bm{FG}$ method is applied to determine the 
normal modes. 
In the 
$[N]$ and $[N-1,1]$ sectors, the normal coordinates
have mixed radial and angular behavior.
The $[N-2,2]$ normal modes are purely angular due to the absence of
$\bar{\bm{r}}'$ symmetry coordinates. 
The amount of radial/angular mixing is dependent on the 
first-order Hamiltonian terms. 

\smallskip
\noindent{\bf{\small{ 4) The energy. -}}} The energy through first-order
in $\delta=1/D$
% (see Eq.~(\ref{eq:E1}))
\cite{FGpaper,loeser}:
\begin{equation}
\overline{E} = \overline{E}_{\infty} + \delta \Biggl[
\sum_{\renewcommand{\arraystretch}{0}
\begin{array}[t]{r@{}l@{}c@{}l@{}l} \scriptstyle \mu = \{
  & \scriptstyle \bm{0}^\pm,\hspace{0.5ex}
  & \scriptstyle \bm{1}^\pm & , %& \\
  &  \,\scriptstyle \bm{2}   \scriptstyle  \}
            \end{array}
            \renewcommand{\arraystretch}{1} }
%\hspace{-0.50em} \sum_{\mathsf{n}_{\mu}=0}^\infty
(n_{\mu}+\frac{1}{2} d_{\mu})
\bar{\omega}_{\mu} \, + \, v_o \Biggr] \,, \label{eq:E1}
\end{equation}
\noindent has the form of a harmonic energy in terms of the normal mode
frequencies, where  $\bar{E}_\infty$ denotes the energy at the minimum 
of $\bar{V}_{\mathtt{eff}}|_\infty$,
%(\bar{r}_\infty)$ 
 $\mu$ labels the five types of normal modes
%, ${\bf 0}^-$\,, ${\bf0}^+$\,, ${\bf 1}^-$\,, ${\bf 1}^+$\,, and ${\bf 2}$\,,
 (irrespective of the value of $N$, see Ref.~\cite{FGpaper} and Ref.[15]
in \cite{paperI}),
with $n_{\mu}$ representing the total normal mode quanta 
with frequency $\bar{\omega}_{\mu}$;
 and $v_o$ is a constant (defined in 
Ref.~\cite{FGpaper}, Eq.(125)).
The five roots have
multiplicities:
$d_{{\bf 0}^+} = 1,  d_{{\bf 0}^-} = 1,
d_{{\bf 1}^+} = N-1,  d_{{\bf 1}^-} = N-1,
d_{{\bf 2}} = N(N-3)/2$.

\smallskip
\noindent{\bf{\small{ 5) The Pauli principle. -}}}
Eq.~(\ref{eq:E1})  defines the ground state/zero-point energy
as well as the  excited state spectrum by assigning
normal mode quantum numbers consistent with the Pauli principle.
The allowed assignments 
are determined by finding a correspondence between
the normal mode states 
$|n_{{\bf 0}^+},n_{{\bf 0}^-},n_{{\bf 1}^+},n_{{\bf 1}^-},n_{\bf 2}>$ and the
non-interacting states of the three dimensional harmonic oscillator 
$(V_{\mathtt{conf}}(r_i)=\frac{1}{2}m\omega_{ho}^2{r_i}^2)$
 whose restrictions due to antisymmetry are well known.
 In the double limit,
  $D\to\infty$, $\omega_{ho}\to\infty$,  both
 representations are valid, so the quantum numbers can be related.
 The radial and angular
characters separate resulting in
two conditions\cite{harmoniumpra,prl}:
\begin{equation} \renewcommand{\arraystretch}{1} 
\label{eq:quanta}
2 n_{{\bf 0}^-} + 2 n_{{\bf 1}^-} =   \sum_{i=1}^N 2 \nu_i \, ,\,\,\,
2 n_{{\bf 0}^+} + 2 n_{{\bf 1}^+} + 2 n_{\bf 2} = \sum_{i=1}^N  l_i  \,,
\renewcommand{\arraystretch}{1}
\end{equation}

\noindent with $\nu_i$ and $l_i$ the radial and 
 orbital angular momentum quantum numbers respectively of the three dimensional 
oscillator,  and  $n_i = 2\nu_i + l_i$ the ith particle 
energy level quanta defined by: 
$E=\sum_{i=1}^N\left[n_i  +\frac{3}{2}\right] \hbar\omega_{ho} =
\sum_{i=1}^N \left[(2\nu_i + l_i) +\frac{3}{2}\right] \hbar\omega_{ho}$.
This strategy is analogous to Landau's use of the non-interacting system  
in Fermi liquid theory to set up the correct
Fermi statistics as interactions evolve adiabatically\cite{landau}.  

%For ultracold temperatures near $T=0$, 
For ultracold systems, only the lowest angular and radial 
modes will be occupied i.e. phonon, $n_{\bf 2}$, 
and  single-particle radial excitation modes, $n_{{\bf 1}^-}$,
 yielding:
\begin{equation} \renewcommand{\arraystretch}{1} 
\label{eq:quanta3}
2 n_{{\bf 1}^-}  =  \sum_{i=1}^N 2 \nu_i ,\,\,\,\,\,\,\,
2 n_{\bf 2} = \sum_{i=1}^N  l_i \,. 
\renewcommand{\arraystretch}{1}
\end{equation}

Near $T=0$, the system energy becomes too small to excite a radial mode, so
only phonon modes are excited creating a gap that supports superfluid
behavior.

\medskip

\section{Zero-point energy considerations}

\smallskip
\noindent{\bf{\small{ 1) The ground state energy. -}}}
The SPT ground state energy that will be used to analyze the
contributions from the uncertainty principle and the Pauli principle has
been previously calculated and compared to results in the
literature both at unitarity\cite{prl} and across the BCS to unitarity
transition\cite{jlow}. In Fig.~\ref{fig:eight} in the Appendix, we reproduce a figure from
Ref.~\cite{prl} which shows
the excellent agreement at unitarity with benchmark Monte Carlo calculations
reported for values of $N \leq 30$\cite{carlson}.
%Above $N=30$
%I would expect the accuracy of the SPT calculations to decline since
%higher order terms may be needed.
Fig.~\ref{fig:nine} in the Appendix is
taken from Ref.~\cite{jlow} and shows energies across the BCS to unitarity
transition with similarly good
agreement of the SPT results with both experiment and theory.

The formula for the energy, Eq.~\ref{eq:E1}, shows explicitly the contribution
of the Pauli principle in the quanta,
$n_{{\bf 0}^+},n_{{\bf 0}^-},n_{{\bf 1}^+},n_{{\bf 1}^-},n_{\bf 2}$ included in the
sum over the five frequencies, $\bar{\omega}_{\mu}$. For temperatures near
$T=0$, only phonon modes are occupied so only $n_{\bf 2}$ is nonzero. Its value
is obtained as explained above using an adiabatic transition from the
independent particle case for which the restrictions of the Pauli principle
are well known to the interacting case.  Knowing the value of  $n_{\bf 2}$,
the contribution of the Pauli principle can be subtracted from the total
energy to leave a non-Pauli contribution due only to the uncertainty
principle, i.e., the lowest
energy that would be possible if the particles could all reside in
the same lowest state.  The contributing terms include $\bar{E}_\infty$,
the energy at the minimum of the effective potential which includes
effects from the trap as well as from the interactions of the $N$ fermions,
as well as the sum over the multiplicities for each frequency and a
correction term, $v_o$
.
(For the formula for $\bar{E}_\infty$ see Eq. 116 in Ref.~\cite{FGpaper}.)

\smallskip
\noindent{\bf{\small{ 2) The Fermi sea. -}}} While the implications of
the Heisenberg
uncertainty principle apply equally
to bosons and fermions, systems of bosons are not subject to the
Pauli principle.
Thus once a lowest single particle state is determined that satisfies the
Heisenberg uncertainty principle for one of the bosons, all of the bosons
can occupy this state. For fermions, only one fermion (or two fermions of
opposite spin) can occupy the lowest spatial state
that satisfies the uncertainty principle. This results in the concept of
a Fermi sea where degenerate fermions fill all the lowest allowed energy
levels according to the Pauli principle.
From an independent particle view, degenerate Fermi systems have all the
lowest energy states filled, with a ``Fermi surface'' dividing the filled
from the unfilled levels. This ``sea'' of fermions in energy space is
defined by the Fermi energy, the largest occupied energy in the system.
The Fermi sea
Pauli-blocks states below the Fermi energy, thus the Pauli principle
controls the behavior of these degenerate
systems, suppressing processes that require an unfilled level in the
Fermi sea. Such behavior has been observed in
the lab\cite{sanner,deb,margalit}. 
Degeneracy pressure, a striking feature of these systems,
is due to the Pauli principle requiring the occupation of high energy levels
and is responsible for the stability of
degenerate Fermi systems and thus their prevalence in our universe. 

From a collective viewpoint, the concept of a Fermi sea is defined 
in terms of the lowest occupied phonon normal mode allowed by the Pauli
principle.  The occupations of the states in the independent particle
picture dictate the restriction on the minimum number of quanta 
required in the collective motion of the ensemble. For ultracold
systems at $T=0$, only phonon modes are occupied so the Fermi sea 
of occupied independent energy states becomes an energy minimum of the phonon 
collective mode,
with lower energy phonon modes unoccupied, i.e. Pauli-blocked from occupation.
From the independent particle view, the Fermi energy is the energy of the
highest occupied independent energy state, while in the collective view,
the Fermi energy
is the energy of the lowest occupied phonon mode. The quantum numbers in these
two regimes are discussed in the Supplemental Information, Section I.E.
in Ref.~\cite{EPL}.
\smallskip

\noindent{\bf{\small{ 3) The evolution from independent particles to the 
strongly correlated unitary regime. -}}} It is helpful to imagine
the evolution from the independent particle picture
to the strongly
interacting regime at unitarity.  This transition
has been studied previously using the SPT approach for the five
normal mode frequencies
as a function of both particle
number, $N$, and interaction strength\cite{prafreq}.
As the particles begin to interact,
the Hamiltonian governing the behavior of the particles now includes this new
interaction potential term. The independent particles that were
responding only to the trap begin to interact with every other particle
in this $N$-body ensemble evolving into highly-correlated, collective behavior.
In the SPT formalism, these interactions
combine with the harmonic trap
term to produce first-order solutions 
which are normal modes of motion.  The familiar evenly-spaced
three dimensional harmonic
spectrum of the trapped independent particles with frequency $\omega_{ho}$
evolves to a new set of 
spectra that reflect the interacting particles. These new spectra are composed
of the normal mode frequencies, five different frequencies that correspond
to the five irreducible representations underwhich the normal modes
transform.  In Fig.~\ref{fig:ten} in the Appendix, I reproduce three figures from
Ref.~\cite{prafreq} showing the evolution of these five frequencies across the
BCS to unitarity transition for three increasing ensemble sizes.
These five frequencies define five separate,
evenly-spaced spectra. The lowest frequency, $\omega_2$, belonging to the
$[N-2,2]$ irreducible representation corresponds to phonon or compressional
motion which is a completely angular mode that transitions adiabatically
to angular states of the independent particle regime.
At temperatures near absolute zero,
these compressional modes are the only ones occupied with the actual
occupation dictated by the Pauli principle and the degeneracies of this
three dimensional many-body system.  The particles in a phonon mode are
moving in sync with a single frequency and phase, and thus this $N$-body
wave function has a large spatial expanse. Near $T=0$, the individual
particles are no longer
identifiable due to their large overlapping
de Broglie wavelenths yielding a large uncertainty in the position of each
particle. Due to this large uncertainty in position, the average
momentum of the $N$ particles
is allowed by the uncertainty principle to
condense toward a single, very small, absolute value, $|k|$.
As the particles move back and forth in lockstep, from $+k$ to $-k$, the
uncertainty in the momentum is squeezed to $2|k|$.

It is instructive to compare the different
contributions from the uncertainty principle alone 
and from the Pauli principle to the zero-point energy in these two regimes: the
independent particle regime and the strongly interacting unitarity regime.
This evolution can also be viewed as evolving from the weakly interacting
BCS regime with infinitesimally small particle interactions, i.e. with
the particles responding essentially only to the trap, to the
strongly-interacting, strongly-correlated unitary regime.
This is achieved in the laboratory in cold fermion
experiments using a Feshbach resonance to tune the particle interactions.
 In the following discussions, I will assume a spin-balanced
 ensemble of identical fermions as is typical in these cold atom experiments.

\begin{figure}
\includegraphics[scale=1.0]{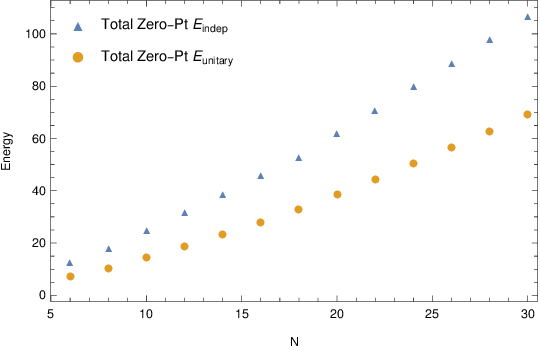}
\renewcommand{\baselinestretch}{1.0}
\caption{Comparison of the SPT zero-point/ground state energies for the independent particle case and the unitarity regime as a function 
  of $N$ in units of $\hbar\omega_{ho}$.}
\label{fig:one}
\end{figure}

Fig.~\ref{fig:one} compares the total zero-point energies in these two
regimes as a function of $N$ for $5 \le N \le 30$.  As expected from the
difference in the spacing of the
spectra for each regime, i.e. $\omega_2  \ll \omega_{ho}$, the independent
particle zero-point energies for an $N$-body system of identical fermions
in an harmonic trap are significantly higher than the zero-point energies for the strongly
interacting unitary regime with the difference increasing
as $N$ increases. (This difference can also be verified in Fig.~\Ref{fig:nine}
in the Appendix which shows the ground state/zero-point energy
decreasing across the BCS to
unitarity transition.)  Clearly, the effect of the Pauli principle for
these two systems of identical fermions is greater for the system with the
more widely spaced spectrum of energy levels (independent particles with
$\omega_{ho}$) that become successively
occupied as the
fermions fill all the lower levels.

\begin{figure}
\includegraphics[scale=1.0]{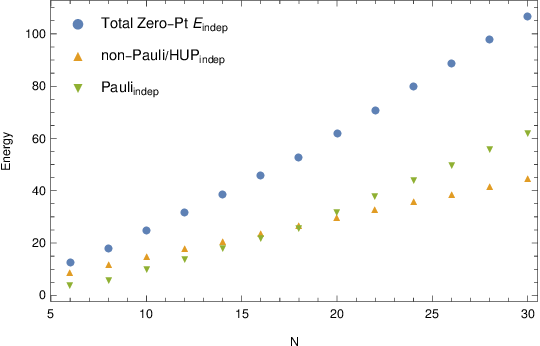}
\renewcommand{\baselinestretch}{1.0}
\caption{The independent particle ground state/zero-point
  energies with the separate contributions
  from the Pauli principle  and the non-Pauli terms as a function 
  of $N$ in units of $\hbar\omega_{ho}$.}
\label{fig:two}
\end{figure}

In Fig.~\ref{fig:two}, the contribution from the Pauli principle is separated
out from the total zero-point energy for the independent particles
leaving the contribution due only to the uncertainty principle.
As one can see for this
regime, the effect of the Pauli principle on the zero point energy starts out
below the uncertainty principle (non-Pauli) contribution, but quickly becomes the larger
contribution with a steeper slope as $N$ increases.

\begin{figure}
\includegraphics[scale=1.0]{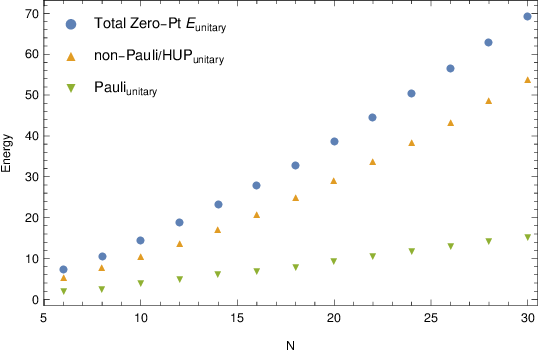}
\renewcommand{\baselinestretch}{1.0}
\caption{The SPT ground state/zero-point energies at unitarity with the contributions from the Pauli and the non-Pauli terms
  as a function of $N$ in units of $\hbar\omega_{ho}$.}
\label{fig:three}
\end{figure}

Fig.~\ref{fig:three} shows a similar breakdown for the zero-point energy in the
strongly interacting unitary regime.  Here the small spacing ($\omega_2$)
of the spectrum
results in a much smaller contribution from the Pauli principle.  The
remaining contribution from the uncertainty principle makes
up the overwhelming majority of the total zero-point energy.

\begin{figure}
\includegraphics[scale=1.0]{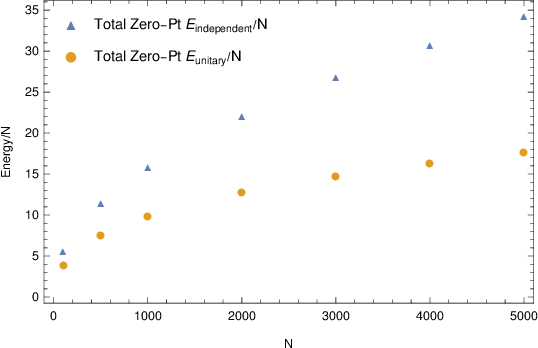}
\renewcommand{\baselinestretch}{1.0}
\caption{Comparison of the SPT zero-point energies per particle
  for the independent particle case
  and the unitarity regime as a function 
  of $N$ for large values of $N$ in units of $\hbar\omega_{ho}$.}
\label{fig:four}
\end{figure}

\begin{figure}
\includegraphics[scale=1.0]{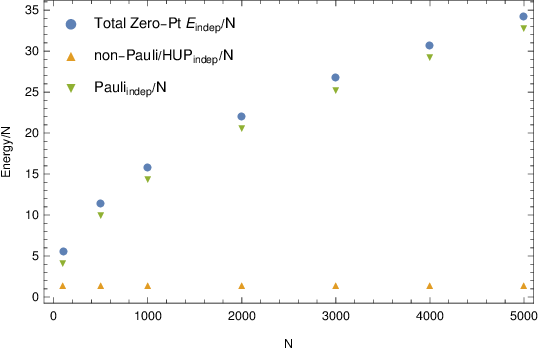}
\renewcommand{\baselinestretch}{1.0}
\caption{The independent particle ground state/zero-point energies per particle
  with the contributions
  from the Pauli principle and the non-Pauli terms as a function 
  of $N$ in units of $\hbar\omega_{ho}$.}
\label{fig:five}
\end{figure}

\begin{figure}
\includegraphics[scale=1.0]{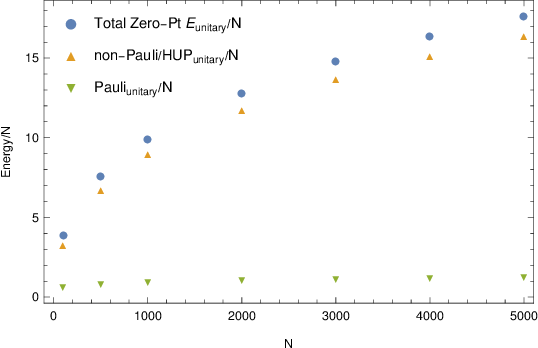}
\renewcommand{\baselinestretch}{1.0}
\caption{The SPT ground state/zero-point energies per particle
  at unitarity with the contributions
  from the Pauli principle and the non-Pauli terms as a function 
  of $N$ in units of $\hbar\omega_{ho}$.}
\label{fig:six}
\end{figure}

Figs.~\ref{fig:four}-\ref{fig:six} show contributions \textit {per particle} for
much higher values of $N$,
$N \le 5000$.
Fig.~\ref{fig:four} compares
the total zero-point energies per particle in these two regimes.
As expected the independent
particle zero-point energies 
are significantly higher per particle than the zero-point energies
per particle for the strongly
interacting unitary regime with the difference increasing
as $N$ increases. 

Analyzing the different contributions in the independent particle regime for
these larger values of $N$,
Fig.~\ref{fig:five} shows that
the Pauli principle is responsible for the overwhelming amount of the total
zero-point energy, tracking close to the total
while the contribution from the uncertainty principle per particle
stays the same for
all $N$ at $3/2$ $\hbar\omega_{ho}$ as expected.

The opposite is true for the interacting regime at unitarity as shown
in Fig.~\Ref{fig:six}.  The contribution per particle
from the Pauli principle effect remains quite small, increasing very
slightly with $N$,
leaving the majority of the zero-point energy per particle contributed by the
non-Pauli terms which are due to the uncertainty principle and closely track
the total zero-point energy.

%Note the range of these energies is smaller than the range of the independent
%particles due to the much smaller spacing of the unitary regime spectrum.

%

Comparing the two graphs, Figs.~\Ref{fig:five} and ~\Ref{fig:six}, (note the
different scales on the vertical axis), the contributions from the Pauli
principle and the uncertainty principle,
i.e. the non-Pauli terms, seem to trade places. In the unitary regime
with interacting particles, we see that the non-Pauli contributions
per particle
to the zero-point energy are larger and increase with $N$,
while the Pauli contributions are
much smaller compared to the independent particle regime.
In contrast, the Pauli principle
dominates in the independent particle
regime, while the non-Pauli uncertainty principle contributions are constant
at $3\hbar\omega/2$ as expected.  This change, of course, takes place
across the BCS to unitarity transition and is due to the increase in
inter-particle interactions that occurs  as
strong correlations set up
collective behavior. These interactions
increase the contribution to the zero-point energy from the
uncertainty principle alone and suppress the contributions from the
Pauli principle as the spectrum is squeezed,
 in effect exchanging the importance of the effects from
these two fundamental principles.
Due to the larger reduction in the Pauli
contribution, the net effect is a lower zero-point energy
for the interacting system as confirmed experimentally.

\smallskip

\section{Discussion}
The difference in the zero-point energy between the independent
particle regime and the strongly interacting unitary regime near $T=0$
is due to several factors that work in concert.
As noted above, I will assume a spin-balanced
ensemble of identical fermions as are used in most ultracold
experiments with atomic fermions.

\smallskip
\noindent{\bf{\small{ 1) Squeezing the Heisenberg uncertainty principle. -}}}
%\noindent{\bf{\small{ a) The independent particle regime. -}}}
In the independent particle regime where the particles are
responding  only
to a harmonic trap, the spectrum is spaced by $\hbar\omega_{ho}$ and
the uncertainty principle is saturated for the lowest
state which is occupied by a pair of fermions with equal
uncertainties in position and momentum. As higher and
higher states fill with fermions, the Pauli principle has a larger and larger
effect dominating the zero-point energy as $N$ increases.

%\noindent{\bf{\small{ b) A squeezed ground state in the strongly-interacting
%unitary regime. -}}}
In contrast, the interacting unitary regime satisfies the
uncertainty principle with a large uncertainty in position 
and a very small uncertainty in the momentum.  To achieve such a small
value of the
momentum requires  the expansion of the uncertainty in position space 
so the Heisenberg uncertainty
principle, whose lower bound was saturated in the lowest state in the
independent particle regime, continues to be satisfied.  This is achieved by the emergence of highly-correlated, collective motion of the fermions near $T=0$ and the overlapping of
their de Broglie wavelengths yielding
a large uncertainty in the position since the particles  can no longer be
distinguished individually. The
phonon normal mode state near $T=0$ with all the particles moving in lockstep
produces this dynamic.
The ground state energy of this
system has been calculated through first order in the SPT perturbation series
using Eq.~\ref{eq:E1} which shows several terms contributing to the
zero-point energy from the uncertainty principle
as well as a contribution from the Pauli principle
from the term that sums over the normal mode quantum numbers
$n_{{\bf 0}^+},n_{{\bf 0}^-},n_{{\bf 1}^+},n_{{\bf 1}^-},n_{\bf 2}$. This ground
state energy agrees well with both experiment and other theoretical work.
See Figs.~\ref{fig:eight} and ~\ref{fig:nine} in the Appendix.
(In Eq.~\ref{eq:E1}, the term, $+\frac{1}{2} d_{\mu}\bar{\omega}_{\mu}$ in
the sum over the normal modes:

\begin{equation}
{\sum_{\mu = {\bm{0}^\pm, \bm{1}^\pm, \bm{2}}}
  {(n_{\mu}+\frac{1}{2} d_{\mu})} \bar{\omega}_{\mu}}
\end{equation}

\noindent appears at first glance to
be the zero-point energy
of this normal mode system.  However, the perturbation derivation for
the ground state energy includes a zeroth order term, $E_{\infty}$ and a
correction term
$ v_o$ that must be included to give the correct contribution.)

%\noindent{\bf{\small{ c) A squeezed spectrum. -}}}
The squeezed value of the momentum yields not only a very low energy ground
state, but an entire spectrum of squeezed states for these compressional
phonon modes.
The squeezed spacing, $\hbar\omega_2$, is much smaller than the spacing
of the independent particle spectrum, $\hbar\omega_{ho}$.

\smallskip

\noindent{\bf{\small{ 2) Suppressing the Pauli principle. -}}}
Since the value of $\omega_2$ is so small,  $\omega_2  \ll \omega_{ho}$,
the Pauli contribution,
$n_2\hbar\omega_2$ where $n_2$ is the quantum number of the
lowest allowed phonon mode, will be quite small for the typical condensate
sizes in experiments compared to the rest of
the zero-point energy.  This keeps the zero-point
energy low, well below the next
higher normal mode state, thus setting up a gap that supports superfluid
behavior.

\begin{figure}
\includegraphics[scale=1.0]{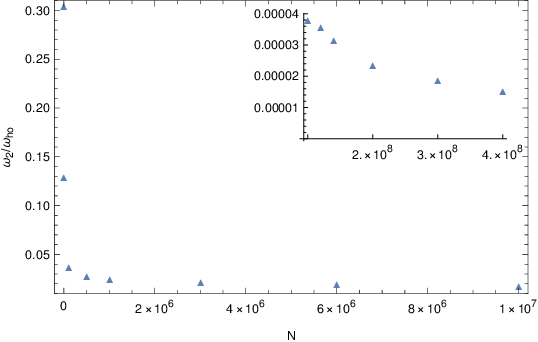}
\renewcommand{\baselinestretch}{1.0}
\caption{The SPT phonon normal mode frequency, $\omega_2$ as a function
  of particle number $N$. The inset extends to very low values of
$\omega_2$ as $N$ reaches values of $10^8$}
\label{fig:seven}
\end{figure}

As $N$ increases, i.e. as
the ensemble size increases, the value of $\omega_2$ decreases as shown in
Fig.~\Ref{fig:seven} so the energy levels being filled according to the
Pauli principle are merging closer and closer in energy. It is interesting
to consider the large $N$ limit where $\omega_2$ limits to
zero (see Appendix G in Ref.~\cite{prafreq}) so the energy levels
are infinitesimally close both to each other and to the lowest phonon level.
This means that the fermions can obey the Pauli principle while
occupying states essentially in the same energy level and infinitesimally
close to the lowest allowed phonon mode, the mode
that would be
occupied if the particles didn't need to obey the Pauli principle. Unlike
degenerate, independent fermions in a harmonic well which fill higher
energy states
with frequencies increasing by integer multiples of $\hbar\omega_{ho}$,
the phonon collective motion at these low temperatures allows the fermions
to occupy energy states that are essentially in the same energy level, i.e. 
behaving in this regard as
condensed bosons would, thus circumventing this effect of the
Pauli principle completely. (Of course, the fermions in these degenerate
or nearly degenerate states are spread out in different wave functions as
required by the Pauli principle in contrast to the case of $N$ bosons which
can all be in the same ground state wave function
resulting in different $N$-body
wave functions for these two systems
which could affect other properties and the overall dynamics.)

Fig.~\Ref{fig:seven} shows the very small SPT values for $\omega_2$ for large
values of $N$ demonstrating the  closeness of the different phonon levels
as this spectrum is squeezed. Note that the decrease in $\omega_2$ 
implies a corresponding decrease in the momentum for these particles
which are moving with the same frequency and displacement.
(See Section C in the Supplementary Information in Ref.~\cite{EPL}.)
Correspondingly,
there
is an increase in the size of the system as $N$ increases which implies an
increase in the uncertainty in the position of each particle
as required by the uncertainty principle. Thus, Fig.~\Ref{fig:seven} verifies
the expected response of complimentary variables obeying the Heisenberg
uncertainty principle at the lower bound of this fundamental principle.

\smallskip
\noindent{\bf{\small{ 3) Manifesting the wave nature of matter. -}}}
In item 2) above, I describe an interesting phenomenon in which
this effect of the Pauli principle is 
suppressed and eventually becomes completely circumvented for $N$ very
large resulting in an ensemble
of identical fermions behaving like identical bosons as they occupy
merged energy levels.
This suppression of the Pauli principle is due to the
squeezed spacing of the phonon spectrum and facilitates
the creation of a gap between the fermions in this lowest normal phonon
mode and the next higher normal mode. To achieve this extremely small
spacing required a large increase in the uncertainty of the position of
each particle. This, at first, might seem difficult to imagine at
these ultralow temperatures when the particles are moving so slowly. The
naive view would be that the particles' position uncertainties at these cold
temperatures and slow movements would also be quite small as they
make tiny compressional movements.  However, as we have seen, the
wave nature of the particles becomes manifest due to both the large
de Broglie wavelengths near $T=0$, but also due to the collective
motion of the particles controlling the size and phase of the
movements. In the large
$N$ limit, the fermions begin to
occupy infinitesimally close energy levels as the energy states
merge into the lowest state and a coherent wave function emerges
manifesting macroscopically the wave nature of matter.

\smallskip
\noindent{\bf{\small{ 4) Engineering the suppression of the Pauli principle. -}}}
One can imagine a second possible way to achieve a suppression of this effect
of the Pauli principle which does not require these ultralow temperatures.
This strategy would involve engineering
a system in which the
degeneracy (or near degeneracy) of the lowest state(s) is greatly enlarged so
that many fermions
can reside at the same (or very close) energy as bosons do.
This would require the
construction of a system with this degeneracy built into the confinement
of the particles or the particle interactions or both
in order to produce
highly degenerate states.  In the current case, the degenercies are constant
as the system transitions adiabatically from the independent particle regime,
for which the degeneracies are well known,
to the strongly interacting unitary regime
(See Section IV in Ref.~\cite{harmoniumpra}) so this proposed
strategy of circumventing
the Pauli principle is not relevant to
the present study, but could be an interesting possibility to construct
useful gaps in other spectrums of interest.

Note, that these two strategies to suppress the Pauli principle are
distinct from the
phenomenon of Pauli blocking where the Pauli principle suppresses other
processes that need an unfilled state
below the Fermi surface. For a degenerate
fermionic ensemble these interactions are called 'Fermi blocked'
by the filled Fermi sea\cite{sanner,deb,margalit,jannin}.

\medskip

\medskip

\section{Conclusions}
Understanding the concept of zero-point energy has been a longstanding
goal in multiple fields of physics since the advent of quantum
mechanics. This fundamental concept representing perhaps the simplest
manifestation of the wave nature of matter affects diverse
phenomena in our universe and is an active area of research in many
areas including the rate of cosmic expansion\cite{maggiore},
a possible source of dark energy\cite{maggiore},
collective fluctuations of complex molecules\cite{richard} and
ultracold atoms\cite{diener,salasnich2},
nanocrystals\cite{duan}, optomechanical systems\cite{lecocq,chegnizadeh},
and quantum entanglement\cite{rodriguez} among other studies.

In this paper, I have investigated the zero-point energy of an ultracold
Fermi gas using the formula for the ground state energy from the
SPT formalism to separate out the effect of the Pauli principle
from the role of the Heisenberg uncertainty principle.
The Pauli principle's role in collective motion was previously used to
reimagine and reinterpret some of the seminal concepts of superfluidity.
Now this collective behavior is studied to understand how the wave-like nature
of identical fermions at ultracold temperatures becomes manifest
and results in a squeezed spectrum of closely spaced energy levels
that supports the emergence of superfluid behavior.
The perturbation
method employed to analyze these results, SPT, is solved exactly at first
order using
group theory to yield normal modes obtaining
a detailed microscopic 
view of the underlying dynamics.

The factors that combine to produce the dynamics to support
superfluid behavior include: 1) many-body
interactions giving rise to strongly correlated, lockstep, collective motion;
2) ultracold temperatures producing large overlapping de Broglie wave
lengths that create large uncertainties in the position of each fermion;
and finally, 3) the squeezed or condensed value of the momentum of each
particle that allows this effect of the Pauli principle to be strongly
suppressed as the fermions occupy infinitesimally close states near the
lowest energy state.

The effects of these two fundamental principles are well known
in the independent particle regime, but, as we have seen, they
produce drastically different effects as the
particles begin to interact in this ultracold regime.
This new dynamic allows the N-body
system of identical fermions to converge toward condensed energy levels
at this low temperature adopting superfluid
behavior which is approaching closer and closer microscopically to bosonic superfluidity.

This collective perspective previously offered an explanation for universal
behavior in the unitary regime. The absence of interparticle interactions
typically defines a regime of independent particles, but can also be
 the defining characteristic of  a strongly correlated ensemble with
particles moving in lockstep with fixed interparticle
distances, i.e. angular normal modes. 
Both types of ensembles are independent of the microscopic
details despite having very different underlying dynamics.

Analyzing the zero-point energy from a collective perspective has
revealed interesting roles of the Heisenberg uncertainty principle and
the Pauli principle working in concert microscopically
at these ultracold temperatures
to create a
superfluid, a powerful macroscopic manifestation of the wave nature of matter.

%\noindent{Acknowledgments. -}
%I am grateful to the National Science Foundation for financial support
%under Grant No. PHY-2011384.

\section*{Declarations}

\begin{itemize}

\item Funding

\smallskip

\noindent This research did not receive funding.

%\noindent I am grateful to the National Science Foundation for financi%al support
%under Grant No. PHY-2011384.

\bigskip

\item Competing interests

\smallskip

\noindent No, I declare that the author has no competing interests as defined
by Springer, or other interests that might be perceived to influence the
results and/or discussion reported in this paper. 
The author has no relevant financial or non-financial interests to disclose.

\bigskip

\item Compliance with Ethical Standards

\smallskip

\noindent The author has no potential conflicts of interest.

\bigskip

\item Consent for publication

\smallskip

\noindent All of the material is owned by the author and/or no permissions are required. 

\bigskip

\item Data Availability

\smallskip

\noindent The data generated or analysed during this study are available from the
author on reasonable request.

\bigskip

\item Code Availability

\smallskip

\noindent Not applicable 

\bigskip

\item Authors' contributions

\smallskip

\noindent D.K. Watson, as the sole author, was responsible for the design of
this research, the investigation, writing the original draft of the manuscript 
and editing the final draft.

\end{itemize}

\bigskip

%\clearpage

\appendix
\renewcommand{\theequation}{A\arabic{equation}}
\setcounter{equation}{0}
\renewcommand{\thefigure}{A\arabic{figure}}
\setcounter{figure}{0}

%\clearpage

\section{The ground state of an ultracold Fermi gas.}
\label{app:ground state}

\begin{figure}
\includegraphics[scale=.6]{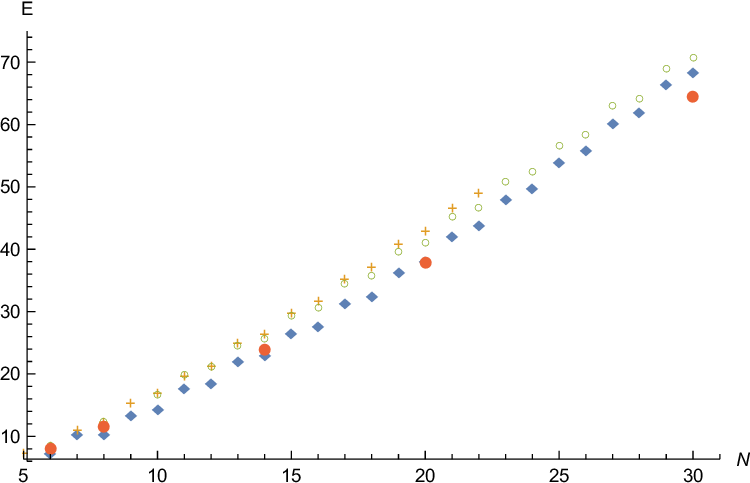}
\renewcommand{\baselinestretch}{.9}
\caption{Ground state energies of the harmonically trapped unitary Fermi gas
(units $\hbar{\omega}_{ho} = 1$). My SPT first-order perturbation
results\cite{prl} 
(filled diamonds) are compared to GFMC (+'s)\cite{chang}, fixed-node 
DMC (open circles) from Ref. \cite{blumestechergreene}, and 
AFMC results (filled circles, $N$=6, 8, 14, 20, 30 only)\cite{carlson}.}
\label{fig:eight}
\end{figure}

The ground state energy of an ultracold Fermi gas has previously been
calculated using the SPT formalism and compared to other theoretical
studies including benchmark Monte Carlo results. Fig.~\ref{fig:eight} below
is taken from Ref.~\cite{prl} and shows the excellent agreement between
SPT results and the benchmark results for values of $N$ reported up to
30\cite{carlson,chang,blumestechergreene}.

\begin{figure}
\includegraphics[scale=0.6]{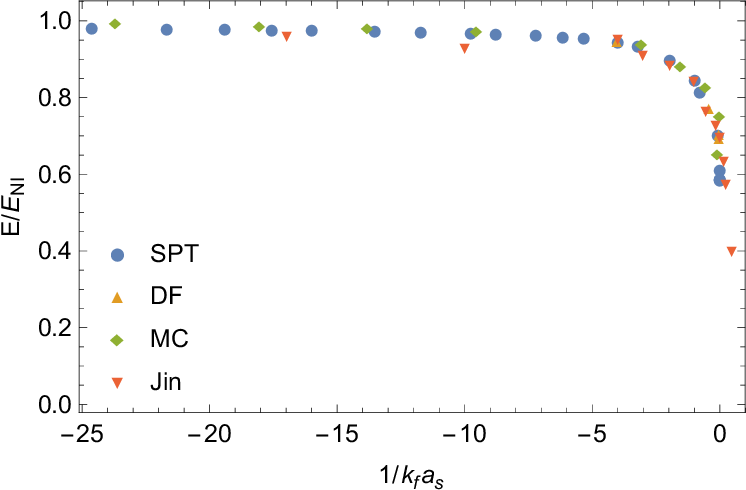}
\renewcommand{\baselinestretch}{0.8}
\caption{Ground state energies from BCS to unitarity as a function of 
$1/k_fa_s$. My SPT results are for $N = 12$\cite{jlow} and are compared to 
experimental \cite{jin3}, density functional (DF)\cite{adhikari1}
and variational Monte Carlo results (MC)\cite{sanchez}.}
\label{fig:nine}
\end{figure}

Fig.~\ref{fig:nine} is taken from Ref.~\cite{jlow} and shows the comparison
of the SPT ground state energies from the BCS regime of weak interactions
to the unitary regime, comparing closely with both experimental
and theoretical work in the literature\cite{jin3,adhikari1,sanchez}.

\begin{figure}
\centering
\begin{subfigure}[h]{0.51\textwidth}
\includegraphics[scale=0.59]{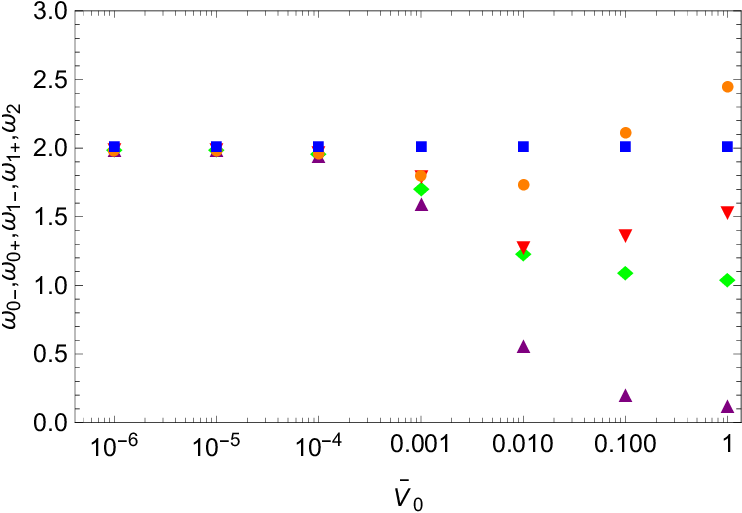}
\subcaption{a. $N=10^3$ fermions.}
\label{fig:trialthirteen}
\end{subfigure}
\begin{subfigure}[h]{0.51\textwidth}
\includegraphics[scale=0.59]{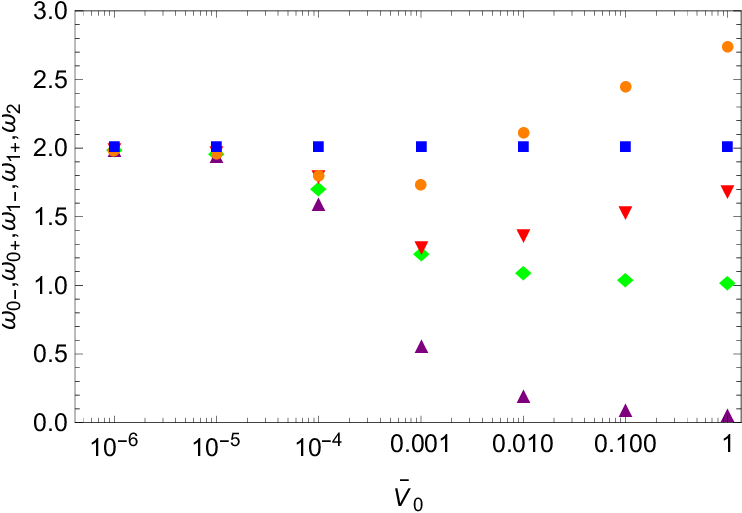}
\subcaption{b. $N=10^4$ fermions.}
\label{fig:trialfourteen}
\end{subfigure}
\begin{subfigure}[h]{0.51\textwidth}
\includegraphics[scale=0.59]{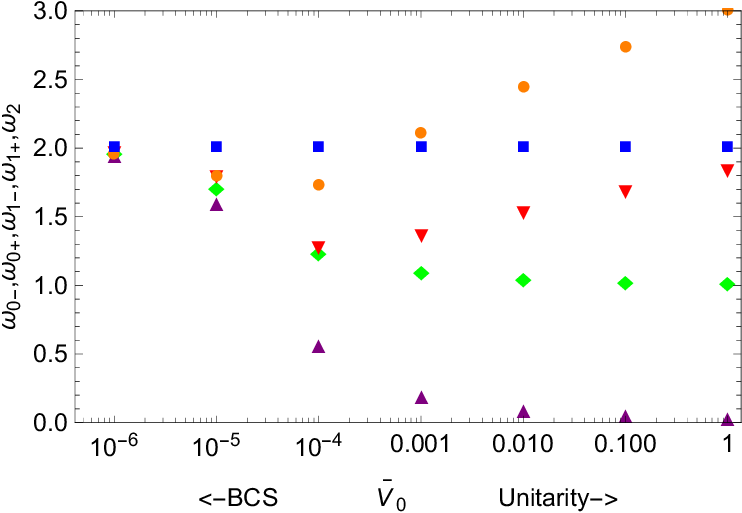}
\subcaption{c. $N=10^5$ fermions.}
\label{fig:trialfifteen}
\end{subfigure}
\caption{SPT frequencies as a function of the interparticle interaction
strength, $\bar{V}_0$, from BCS to unitarity in units of the trap frequency
from Ref~\cite{prafreq}.
($\omega_{{0}^-}$ orange dots; $\,\,\omega_{{0}^+}$ blue squares; 
$\,\,\omega_{{1}^-}$ red 'down' triangles; $\,\,\omega_{{1}^+}$ green diamonds;
$\,\,\omega_{2}$ purple 'up' triangles). Note the log scale on the x axis.}
\label{fig:ten}
\end{figure}

In Fig.~\ref{fig:ten}, I reproduce three figures from Ref.~\cite{prafreq} showing the
behavior of the five normal mode frequencies across the transition from the
non-interacting regime to the strongly-interacting unitary regime for three
different sizes of the ensemble. One can see that the angular frequencies
converge to integer multiples of the trap frequency, $\omega_{ho}$, with the
phonon frequency limiting to 'zero' times $\omega_{ho}$.  See Section VI. D.
in Ref.~\cite{prafreq} for a discussion of the microscopic dynamics underlying this behavior and
Appendix E in Ref.~\cite{prafreq} for the derivation of this limiting behavoir
for the phonon mode.

From these three graphs, one can see that as $N$ increases and thus the spatial
expanse of the wave function increases yielding a larger uncertainty
in the position of each particle, $\omega_2$ decreases
%at lower values of the interaction strength,
 implying a decrease in the 
uncertainty in the momentum as allowed by the uncertainty principle.  The phonon states
filling with fermions merge closer and closer together so the effect of the
Pauli principle continues to be minimized retaining the gap to
higher normal modes which supports superfluidity.

\end{document}